# A Note on the Relationship between Sunspot Numbers and Active Days

J.M. Vaquero[1*], S. Gutiérrez-López[1], and A. Szelecka[1,2]

[1]Departamento de Física, Universidad de Extremadura, Mérida (Badajoz), Spain

[2]Faculty of Physics and Astronomy, University of Zielona Góra, Poland

**Abstract**

We have studied the relationship between three different versions of the sunspot number (Group, International and American sunspot number) and the number of active days (i.e., the number of days with spots on the solar disk). We have detected an approximately linear relationship for low solar activity conditions. However, this relationship for the International sunspot number is very different to the ones obtained with the other versions of the sunspot number. The discordant values correspond to older observations.

**1. Introduction**

The unusual solar activity of the last years (Russell et al., 2010; Schrijver et al., 2011) has caused a growing interest in solar activity over the past centuries and, especially, on the main series that is used in these studies: the sunspot number (SN). However, we encounter a major problem. Two different series of Sunspot Number are available for the last few centuries (Vaquero, 2007; Cliver et al., 2013; Usoskin, 2013). On the one hand, we have the International Sunspot Number (ISN) that is available at daily scale from 1818 to the present and is maintained and updated by the Solar Influence Data Analysis Center (SIDC; Clette, 2011). It is a successor of the original Wolf Sunspot Number (Wolf, 1861). On the other hand, we also have the Group Sunspot Number (GSN), constructed by Hoyt and Schatten (1998) and based on the number of sunspot groups. It is available from 1610-1995 at daily scale but with significant gaps. Additionally, the American Sunspot Number (ASN) is also available but only since 1944 (Hossfield, 2002).

---

[*] Corresponding author (jvaquero@unex.es)

Given the significant differences between the ISN and the GSN in the historical period (mainly before 1880), we think that we need to work with an index of solar activity even easier that the Sunspot Number: the yearly number of active days (AD: days with sunspots reported on the solar disc, usually expressed as percentage of active days per year). In fact, the discovery of the solar cycle by Schwabe (1844) was made possible by a complementary index (inactive days with no spots). Maunder (1922), Harvey and White (1999), Usoskin et al. (2000, 2001), and Kovaltsov et al. (2004) used AD as a reliable indicator of solar activity. Recently, Vaquero et al. (2012) have used the relationship between annual sunspot numbers and AD to detect inconsistencies in low annual GSN. Moreover, Chang (2013) has found that the slope of the linear relationship between monthly ISN and the monthly number of AD is dependent on the solar activity at its maxima.

The aim of this work is to study this relationship between annual sunspot numbers and AD using different versions of Sunspot Number (GSN, ISN and ASN) to detect inconsistencies and differences between them.

**2. Sunspot Numbers vs. Active Days**

We have used the daily values of GSN, ISN, and ASN from 1610, 1818, and 1945 respectively. These data are available at the National Geophysical Data Center and the SIDC (http://ngdc.noaa.gov/ and http://sidc.oma.be/ respectively).

Figure 1 shows the relationships between different annual values of SNs (GSN, ISN, and ASN) and AD. Vaquero et al. (2012) have shown that the relationship between GSN and AD is approximately linear for low values of AD. Moreover, they show that some values of annual GSN does not follow this empirical linear rule. Recently, Vaquero and Trigo (2014) have corrected some of these values consulting the original observations.

Despite the short length of ASN series, the relationship between ASN and AD values is very similar to the relationship that is obtained using the GSN values. However, this relationship is different when one uses ISN values, especially for the lower values of this index. Table 1 list the main parameters of the best linear fit for values of AD<85%, showing clearly the significantly non-zero offset. Note that it is unphysical because SN should be close to zero for very small values of AD.

**3. Discussion**

We wanted to analyze more carefully the difference in the relationship between ISN and AD. To do this, we have divided the original series into parts of three solar cycle length to check if this difference is localized in specific periods or is systematic.

Thus, we have computed for each three solar cycles the relationship between ISN and AD (Figure 2, upper panel) and we have adjusted the data to the best linear fit for values of AD<85%. Table 2 lists the main parameters of these fits.

From theoretical considerations, we can establish the relationship ISN = 0.11 AD, assuming that for low solar activity only one isolated spot is in the sun disc per each active day. ). Note that the most important aspect of the analysis is that the ISN must go to zero when AD approaches zero and a value of ISN>0 for AD=0 is unphysical. Table 2 shows that the most anomalous values with respect to the theoretical values correspond to SC 7-9. This piece of the series is the earliest period with many data prior to the observations by Wolf.

We can repeat this analysis taking into account the history and development of the sunspot number (Clette et al., 2007; Cliver et al., 2011). We are using the intervals 1818-1848, 1849-1875, 1876-1893, 1894-1946, 1947-1980, and 1981-2012. Note that the smallest non-zero ISN value is 11. However, this value can be different due to the calibration constant of the observers. In the interval 1818-1848, the lowest non-zero value is mainly 14, because Wolf increased the values of Schwabe (main observer in this period) by a factor 5/4=1.25. Therefore, the lowest non-zero value is 14. From 1849-1875, the main observer is Wolf and, therefore, the lowest non-zero value is 11 in general because the calibration constant is equal to 1. Between the years 1875 and 1893, the main observer is Wolf but solar observations of his assistants (mostly Wolfer) were also used. After the death of Wolf, from 1894, Wolfer multiplied his counts by 0.6. Therefore, the lowest non-zero value of ISN is 7 in general. Moreover, Svalgaard (2011) detected a jump in the ISN values around 1946 and, since 1980, the modern computation of ISN is made by SIDC.

We have drawn the relationship between ISN and AD (Figure 2, lower panel) for these intervals and we have adjusted the data to the best linear fit for values of AD<85%. Table 2 lists the main parameters of these fits. The result showing clearly a significantly non-zero offset for the first time interval (1818-1848) in agreement with our first analysis. The high values of that ratio for SC 7-9 on Figure 2 (upper panel) for very low AD could be related with the fact that the lowest non-zero values of ISN is 14 and not 7.

**4. Conclusion**

We have studied the relationship between three different versions of the sunspot number (GSN, ISN, and ASN) and AD detecting an approximately linear relationship for conditions of low solar activity. However, the relationship for the ISN is very different to the ones obtained with GSN and ASN. Table 1 lists the main parameters of the best linear fit for values of AD<85%, showing clearly a significantly non-zero offset for ISN. Note that it is unphysical because SN should be close to zero for very small values of AD. We have found that discordant values were derived from older observations (SC 7-9, approximately from 1823 to 1855). These historical values are based primarily on observations by S. H. Schwabe (Arlt et al., 2013) that are reliable observations. However, Wolf used a calibration constant equal 1.25 for the sunspot counts made by Schwabe and, therefore, the lowest non-zero value of ISN is mainly 14 in the interval 1818-1848. Thus, the problem could be related to the calibration constants (Leussu et al., 2013) and the non-linearity of ISN for low values.

**Acknowledgements**

J.M. Vaquero has benefited from the impetus and participation in the Sunspot Number Workshops (http://ssnworkshop.wikia.com/wiki/Home). Support from the Junta de Extremadura (Research Group Grant No. GR10131), Ministerio de Economía y Competitividad of the Spanish Government (AYA2011-25945) and the COST Action ES1005 TOSCA (http://www.tosca-cost.eu) is gratefully acknowledged. We would also like to thank the three anonymous reviewers for their positive and constructive comments.

Table 1. Main parameters of the linear fits to the relationship between annual International, Group and American Sunspot Number and AD (expressed as percentage of active days per year) for values of AD<85%: slope, y-intercept, and correlation coefficient (r).

|  | Slope | y-intercept | r |
|---|---|---|---|
| ISN | 0.14±0.03 | 12±2 | 0.269 |
| GSN | 0.23±0.01 | -0.3±0.6 | 0.655 |
| ASN | 0.29±0.05 | -8±4 | 0.789 |

Table 2. Main parameters of the linear fits to the relationship between annual ISN and AD (expressed as percentage of active days per year) for values of AD<85% taking into account different time intervals: (upper panel) consecutive three solar cycle periods and (lower panel) stages in the history of ISN.

|  | Slope | y-intercept | r |
|---|---|---|---|
| SC 7-9 | 0.15±0.03 | 17±2 | 0.768 |
| SC 10-11 | 0.22±0.08 | 7±5 | 0.653 |
| SC 13-15 | 0.16±0.04 | 8±2 | 0.794 |
| SC 16-18 | 0.13±0.07 | 10±4 | 0.595 |
| SC 19-21 | 0.13±0.25 | 10±18 | 0.251 |
| SC 22-24 | 0.21±0.04 | 4±3 | 0.910 |
|  |  |  |  |
| 1818-1848 | 0.14±0.03 | 18±2 | 0.703 |
| 1849-1875 | 0.18±0.005 | 11±3 | 0.740 |
| 1876-1893 | 0.2±0.1 | 7±6 | 0.353 |
| 1894-1946 | 0.15±0.04 | 9±2 | 0.540 |
| 1947-1980 | 0.13±0.08 | 10±6 | 0.371 |
| 1981-2012 | 0.21±0.04 | 4±3 | 0.808 |

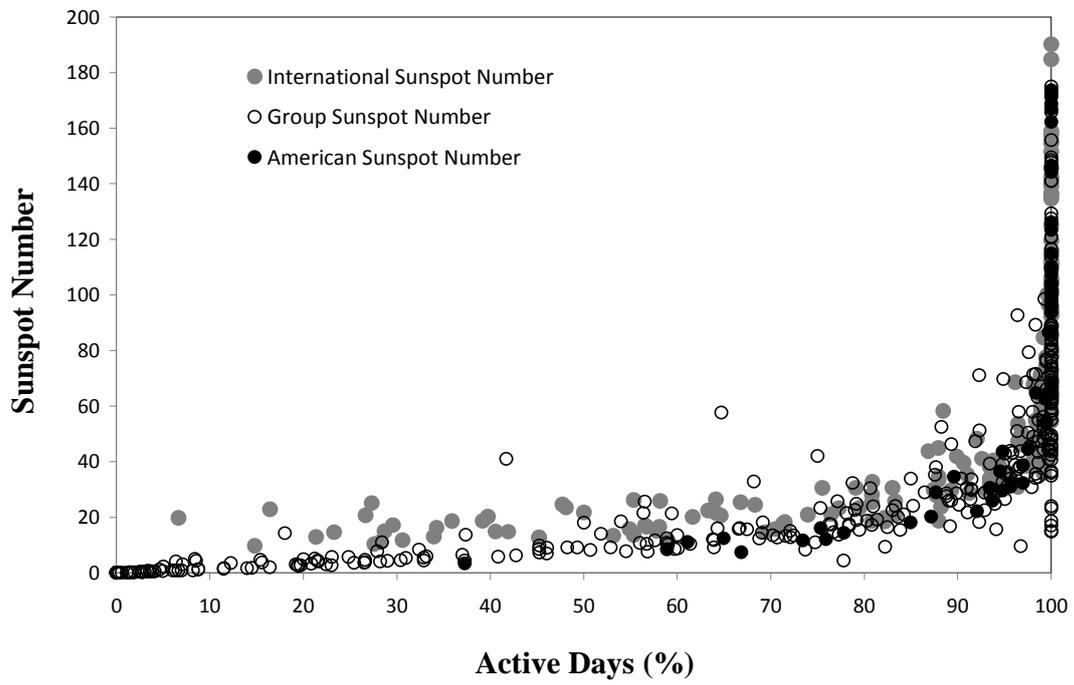

Figure 1. Relationship between the annual International, Group and American Sunspot Number and Active Days (AD, expressed as percentage of active days per year).

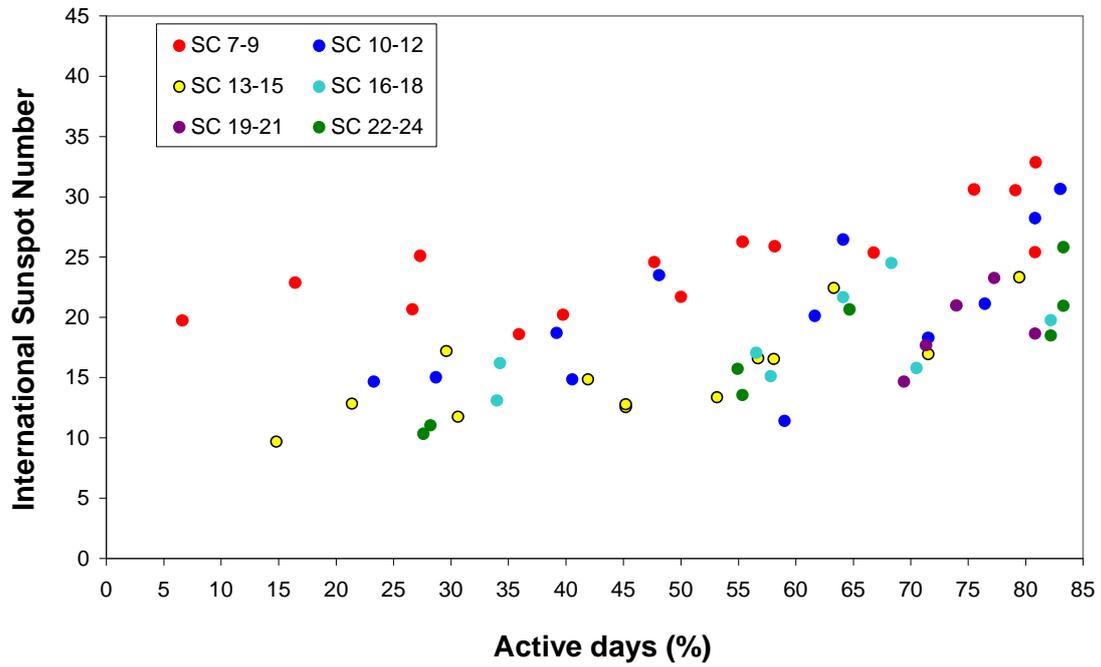

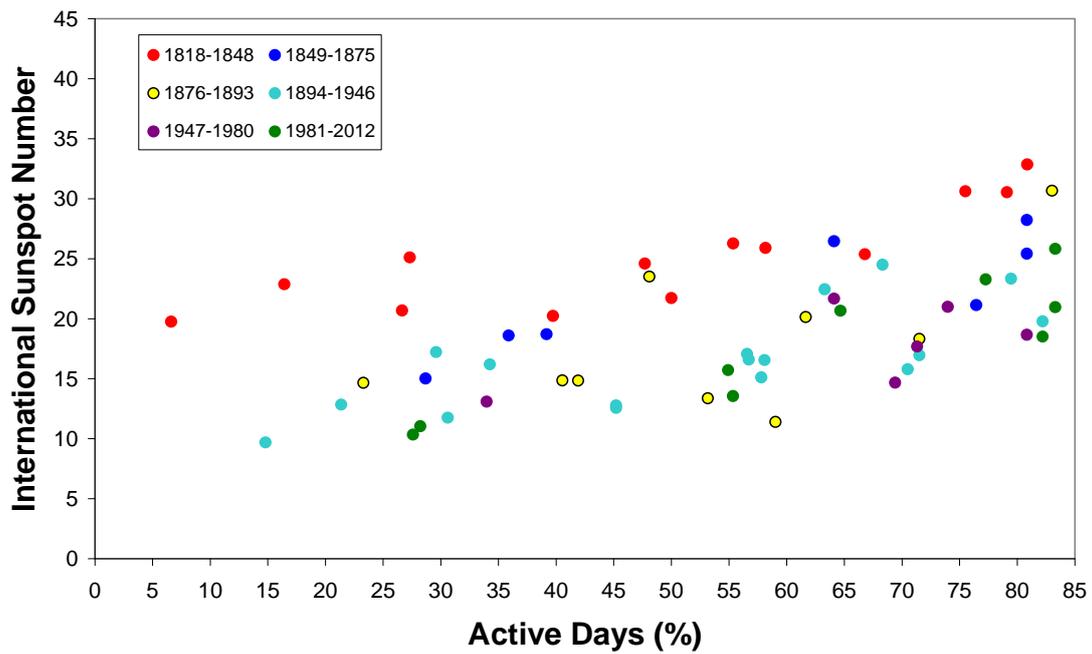

Figure 2. Relationship between the annual International Sunspot Number and Active Days (expressed as percentage of active days per year) for values of AD<85% taking into account different time intervals: (upper panel) for consecutive three solar cycle periods and (lower panel) for stages in the history of ISN.

9